\documentclass[11pt,a4paper]{article}

\usepackage{geometry}
\usepackage[latin1]{inputenc}

\usepackage{comment}
\usepackage{float}
\usepackage{listings}
\usepackage{mathrsfs}
\usepackage{amsmath}
\usepackage{amssymb}
\usepackage{multicol}
\usepackage{float}
\usepackage{makeidx}
\usepackage{layout}
\usepackage{array}
\usepackage{hyperref}
\usepackage{latexsym}
\usepackage{indent first}

\usepackage{subcaption,graphicx}
\usepackage{appendix}
\usepackage{multirow}

\usepackage{color}

\usepackage[sectionbib,square]{natbib}

\usepackage[normalem]{ulem}

\newcommand{\R}{\mathbb{R}}
\newcommand{\X}{\mathbf{X}}
\newcommand{\Y}{\mathbf{Y}}
\newcommand{\F}{\mathcal{F}}
 \newcommand{\C} {\mathcal{C}}
 \renewcommand{\H}{\mathcal{H}}
\newcommand{\Phib}{\mathbf{\Phi}}
\newcommand{\Psib}{\mathbf{\Psi}}

\newcommand{\hs}{\hspace}

\newcommand{\ds}{\displaystyle}

\newcommand{\phib}{\overline{\phi}}
\newcommand{\psib}{\overline{\psi}}
\newcommand{\U}{\overrightarrow{U}}

\DeclareMathOperator{\sech}{sech}

\title{Numerical stability of solitary waves in flows \\ with constant vorticity for the Euler equations}
\author{ Eduardo M. Castro$^{1}$, Marcelo V. Flamarion$^{2}$ and Roberto Ribeiro-Jr$^{3}$}

\date{}

\begin{document}

\maketitle

\begin{center}
{\footnotesize $^1$Graduate Program in Mathematics \\
   UFPR/Federal University of Paran{\' a} \\
   Centro Polit\'ecnico, Jardim das Am\'ericas, Curitiba-PR, Brazil, 81531-980\\
   eduardomdecastro@gmail.com
}\\
\vspace{0.3cm}
{\footnotesize $^2$Unidade Acad{\^ e}mica do Cabo de Santo Agostinho\\
	UFRPE/Rural Federal University of Pernambuco \\
	BR 101 Sul, Cabo de Santo Agostinho-PE, Brazil,  54503-900 \\
marcelo.flamarion@ufrpe.br }

\vspace{0.3cm}
{\footnotesize $^3$Departament of Mathematics\\
   UFPR/Federal University of Paran{\' a} \\
   Centro Polit\'ecnico, Jardim das Am\'ericas, Curitiba-PR, Brazil, 81531-980\\
   robertoribeiro@ufpr.br
}


\end{center}

	
	
	
	
\begin{abstract}

\noindent  The study of the  Euler equations in flows with constant vorticity has piqued the curiosity of a considerable number of researchers over the years. Much  research has been conducted  in this  subject  under the assumption of steady flow. 	In this work, we provide a numerical approach that allows to compute   solitary waves in flows with constant vorticity and analyse  their  stability.  Through a conformal mapping technique,  we compute  solutions of the steady Euler equations, then feed them  as initial data for the time-dependent  Euler equations.  We focus on analysing to what extent the steady solitary waves 
 are stable within the time-dependent framework. Our numerical simulations indicate that   although  it is possible to compute   solitary waves for the steady Euler equations in flows with large values of vorticity,  such waves are not numerically stable for  vorticities with absolute value  much greater than one. Besides, we notice that large waves are unstable even for small values of vorticity. 
  \vspace{10pt}

\noindent {\bf Key words:} Constant vorticity, Solitary water waves, Euler equations, Numerical stability. 
 
\end{abstract}

\section{Introduction}

{The propagation of solitary water  waves on flows with constant vorticity  has piqued the curiosity of a considerable number of mathematicians, engineers and physicists over the years.  Theoretical, numerical and experimental studies have  been conducted on this topic.}





%

The  first rigorous existence  theory of solitary waves to the Euler equations dates back to \cite{Lavrentiev}, \cite{Ter-Krikorov},  \cite{Friedrichs}  and \cite{Beale} for small amplitude waves, and  to  \cite{Toland1, Toland2, Toland3} for large-amplitude solitary waves. All these works considered irrotational flows.  The proof of the existence of solitary waves with vorticity was given by \cite{Hur} (for small amplitude waves) and by \cite{Wheeler} (for large amplitude waves). Both authors used in their works the Dubreil-Jacontin transformation which presupposes the non-existence of stagnation points --  fluid particles with zero velocity in the wave's moving frame.  Only recently, \cite{Kozlov} proved the existence of solitary waves for the Euler equations in flows with constant vorticity allowing  stagnation {points} within the fluid bulk. However, asymptotic works from the 1980s have indicated the existence of solitary waves in flows with stagnation points in the interior \cite{Johnson}.


Asymptotic models for solitary water waves with vorticity were initially studied by  \cite{Benjamin2}, for steady solitary waves, and by \cite{Freeman}, who deduced a KdV type equation from the  Euler equations in the presence of a nonuniform current varying vertically. In these works the authors assumed that the ratio between the water depth and the characteristic wavelength is small (weakly dispersive regime), as well as the ratio between the wave amplitude and the  water depth (weakly linear regime). Without imposing any restrictions on the wave amplitude, \cite{Choi} deduced an asymptotic model for weakly dispersive solitary waves in flows with constant vorticity. More recently, \cite{Guan} studied numerically the structure of the flow beneath  rotational solitary waves. In this work, the author compared the particle trajectories captured by the KdV model with the  ones generated by the  Euler equations. His results show that both models capture practically the same structure inside the fluid (trajectories and stagnation points) when the wave amplitude is small.


One technique commonly used to  deal numerically  with traveling solutions for  the full Euler equations is to consider the steady version of this system.  Among the authors that used this approach we refer to \cite{Da silva e Peregrini}.  Through a conformal mapping that transforms the physical domain into a somewhat anular region, they used a boundary integral  method and numerically  computed solitary waves with vorticity for the steady Euler equations. 
Later, \cite{Vanden-Broeck94} revisited this problem and found numerically  waves with constant vorticity that have overturning profiles, that is, profiles that are not graphs of a function. 
 Furthermore, he showed that there are branches of solutions that do not bifurcate from the trivial shear flow, that is, from the current-induced flow varying linearly with depth to the free surface at rest.

 Regarding the numerical schemes for the time-dependent Euler equations, the technique introduced by \cite{Dyachenko} stands out. Through a time-dependent conformal mapping, the procedure proposed by these authors transforms  the Euler equations into a system of ODEs. This approach has resulted in  numerical methods with good accuracy  as illustrated, for example, by  \cite{Marcelo-Paul-Andre} in the study of the dynamic of waves generated by  current-topography interactions.  We point out that  this conformal mapping  has also been widely applied in the numerical study of the steady Euler equations,   mainly in the investigation of the structures beneath  periodic waves with constant vorticity \cite{DyachenkoHur1, DyachenkoHur2, Ribeiro Jr, Choi2}.
 


  In this  work, we study  numerically the wave stability of solitary waves   in flows with constant vorticity through the Euler equation.  
   More specifically, based on the work of \cite{Ribeiro Jr} for periodic  traveling waves in flows with constant vorticity, we propose a numerical scheme to find solitary waves with constant vorticity for the steady Euler equations. Then, we use these waves as  initial data for the time-dependent  Euler equations.    This approach allows us to analyse whether the steady solutions are indeed  solutions of the time-dependent Euler equations.  In an effort to compute the wave evolution,  we adjust  the numerical method  proposed by \cite{Marcelo-Paul-Andre} in the study of generated waves.  As far as we know there are no study on the wave stability of solitary waves with constant vorticity in the context that we are doing here. Previous studies focus on the wave stability of periodic waves in the sense of Benjamin-Feir instability (\cite{Choi2}) and linear stability (\cite{Kharif}).

   Our results indicate that although there may exist solitary waves for the steady Euler equations for arbitrarily large vorticity values, these waves are not numerically stable. Furthermore, the simulations show that the solitary waves are stable only for small vorticity values.

   For reference, this article is organized  as follows. The governing equations of water waves in flows with constant vorticity are presented in section \ref{Sec_governing_eq}. In section \ref{Sec_steady}, we describe the numerical method  for  solitary waves solution to the steady Euler equations. In  section \ref{time}, we deal with the time-dependent Euler equations. Then, we  move to a compilation of our main results and simulations in  section \ref{Sec_Results} and proceed to our  final considerations.

\section{Governing equations}\label{Sec_governing_eq}

We consider an incompressible flow of an inviscid fluid with constant density ($\rho$) in a  two-dimensional channel with finite depth ($d$) under the force of gravity ($g$), and constant pressure ($P_0$). Besides, we assume that the flow  is in the presence of a linearly sheared current (constant vorticity). Denoting the velocity field in the bulk fluid by $\U(x,y,t)=(u(x,y,t),v(x,y,t))$ and the free surface by ${\zeta}(x, t)$, this free-boundary problem can be described by the Euler equations

\begin{equation}\label{eulers}
    \begin{array}{rclr}
        \U_t+(\U \cdot \nabla ) \U & = & -\ds\frac{\nabla p}{\rho} -g\mathbf{j} & \text{in} \; -d < y < \zeta(x,t),  \\
        \nabla \cdot \U & = & 0 & \text{in} \; -d < y < \zeta(x,t), \\
        p & = & P_0 & \text{at} \; y = \zeta(x,t), \\
        v & = & \zeta_t+u\zeta_x & \text{at} \; y = \zeta(x,t), \\
          v & = & 0 & \text{at} \; y=-d,
    \end{array}
\end{equation}
where $\mathbf{j}$ is the unitary vector $(0,1)$. 

{\color{black} The assumption of constant vorticity enables us to write the velocity field as
 \begin{equation}\label{Ucampo}
 \U=\overrightarrow{U_0}+\nabla \phib,
 \end{equation}
 where 
 $$\overrightarrow{U_0}=(a y+f,0),  \quad f \in \R,$$
 is a  linear shear flow solution of (\ref{eulers})  characterized by the flat surface $\zeta \equiv 0$ and constant vorticity $-a$. Here,   $\phib$ is the velocity  potential of an irrotational perturbation of the shear flow.  The insertion of  (\ref{Ucampo})  in the Euler equations (\ref{eulers}) yields the  set of equations }




\begin{equation}\label{ePotencial}
    \begin{array}{lr}
        \Delta \phib = 0 & \; \text{in} \; -d < y < \zeta(x,t),  \\
        \zeta_t + (a\zeta+f+\phib_x)\zeta_x=\phib_y & \: \text{at} \; y = \zeta(x,t), \\
         \phib_t+\frac{1}{2}(\phib_x^2+\phib_y^2)+(a\zeta+f)\phib_x+g\zeta-a\psib=B(t) & \; \text{at} \; y=\zeta(x,t), \\
         \phib_y = 0 & \; \text{at} \; y=-d,
    \end{array}
\end{equation}
where $\psib$ is a harmonic conjugate of the potential function $\phib$. We make variables non-dimensional via the following transformations:

\begin{equation}\label{rescale}
    \begin{array}{lcr}
         x=dx', & \zeta=d\zeta', & \Omega=\ds\frac{ad}{\sqrt{dg}}, \\ 
         y=dy', & \phib=d\sqrt{dg}\phib', &  p=P_0+\rho gdp', \\
         t=\sqrt{\frac{d}{g}}t', & \psib=d\sqrt{dg}\psib', & F=\ds\frac{f}{\sqrt{dg}}.
    \end{array}
\end{equation}
Dropping the prime notation, this gives us the dimensionless version of  equations (\ref{ePotencial}):

\begin{equation}\label{td}
    \begin{array}{lr}
        \Delta \phib = 0 & \; \text{in} \; -1 < y < \zeta(x,t),  \\
        \zeta_t + (\Omega\zeta+F+\phib_x)\zeta_x=\phib_y & \: \text{at} \; y = \zeta(x,t), \\
        \phib_t+\frac{1}{2}(\phib_x^2+\phib_y^2)+(\Omega\zeta+F)\phib_x+\zeta-\Omega\psib=B(t) & \; \text{at} \; y=\zeta(x,t), \\
        \phib_y = 0 & \; \text{at} \; y=-1,
    \end{array}
\end{equation}
where $-\Omega$ is the dimensionless vorticity and $F$ is the Froude number. 

 For the study of travelling wave solutions it is convenient to eliminate time from the problem by passing to a  moving frame
 $$X=x-ct \quad \mbox{ and } \quad Y=y,$$
 where $c$ is the wave speed,  to be determined {\it a posteriori}.   In this new moving reference frame the wave is stationary and the flow is steady. Taking this new frame of reference into account, equation (\ref{td}) becomes


\begin{equation}\label{travelling}
    \begin{array}{lr}
        \Delta \phib = 0 & \; \text{in} \; -1 < Y < \zeta(X),  \\
        -c\zeta_X + (F+\Omega\zeta+\phib_X)\zeta_X=\phib_Y & \: \text{at} \; Y = \zeta(X), \\
        -c\phib_X+\frac{1}{2}(\phib_X^2+\phib_Y^2)+(\Omega\zeta+F)\phib_X+\zeta-\Omega\psib=B & \; \text{at} \; Y=\zeta(X), \\
         \phib_Y = 0 & \; \text{at} \; Y=-1.
\end{array}
\end{equation}

{\color{black}
We assume that $\zeta(X)$ is a solitary wave whose crest is located at $X=0$ and satisfies} 
\begin{equation}\label{limite}
	\zeta(X) \to 0 \quad \mbox{as} \quad |X| \to \infty. 
\end{equation}
In the following, we present a numerical scheme to compute the solutions of the system (\ref{travelling}) and a numerical scheme to the evolution problem (\ref{td}).




\section{Steady solitary  waves} \label{Sec_steady}

Since $\zeta(X)$ decays to zero as $|X|\to \infty $, we can truncate its infinite domain  to a finite one $[-\lambda/2,\lambda/2]$ with $\lambda >0$, and approximate the boundary conditions by periodic conditions. Then we can solve equations (\ref{travelling})   through  the conformal mapping technique  introduced by \cite{Dyachenko},  that has been widely applied in  free boundary problems (\cite{Choi, Milewski, Ribeiro Jr}.)  	  This strategy consists in using a  conformal mapping  from a strip of length $L$ and width $D$ (canonical domain) onto the flow domain of the solitary wave $\{(X,Y)\in \R^2, -\lambda/2\leq X \leq \lambda/2 \mbox{ and } -1\leq Y \leq \zeta(X)  \}$.  This map is such that  in the canonical domain the free boundary problem  (\ref{travelling}) can be solved numerically by the use a spectral collocation method and Newton's method. 

\subsection{Conformal mapping for steady waves}

\begin{figure}[!ht]
	\centerline{\includegraphics{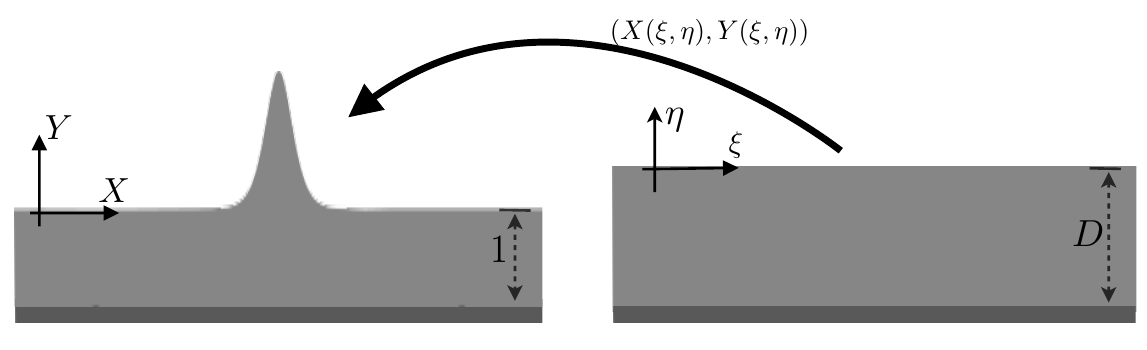}}
	\caption{Illustrative depiction of the conformal mapping. The free surface is flattened out in the canonical domain. }
	\label{Map}
\end{figure}

Consider the conformal transformation 
\begin{equation}
    Z(\xi,\eta)=X(\xi,\eta)+iY(\xi,\eta),
\end{equation}
under which the strip $\{(\xi,\eta)\in \mathbb{R}^2; \;  -L/2 \leq \xi \leq L/2 \mbox{ and } -D\leq \eta \leq 0\}$ is mapped onto the flow domain, as in Figure \ref{Map}. The constant $D$ will be determined so that  both  domain  have the same length. Hence, the mapping does not alter the wavelengths.  Since $Z$ is taken to be conformal, thus analytical, $X$ and $Y$ are actually conjugate harmonic functions, whereas the mapping's Jacobian is given by 
\begin{equation}
   J=X_\xi^2+Y_\xi^2.
\end{equation}

A central characteristic of this mapping is given by the way the boundary curves from each domain are related

\begin{equation}\label{waveprofile}
    \begin{cases}
    Y(\xi,0)=\zeta(X(\xi,0)), \\
    Y(\xi,-D)=-1,
    \end{cases}
\end{equation}
which serves as Dirichlet data for the Laplace's equation for $Y$. {\color{black} By denoting $\mathbf{Y}(\xi)=Y(\xi,0)$ and $\mathbf{X}(\xi)=X(\xi,0)$ the traces of the respective harmonic functions along $\eta = 0$, we have that 


\begin{equation}\label{Ything}
    Y(\xi,\eta)=\mathcal{F}^{-1}\left[\frac{\sinh(k(\eta+D))}{\sinh(kD)} \mathcal{F}(\mathbf{Y})\right]+\frac{(\eta+D)\langle\Y\rangle + \eta}{D}, \quad k\neq 0,
\end{equation}
where $k=k(j)=(\pi/L)j$, for $j \in \mathbb{Z}$,  $\mathcal{F}$ is the Fourier transform in $\xi$-variable given by 

\begin{equation*}
    \mathcal{F}(f(\xi))=\hat{f}(k)=\ds\frac{1}{L}\ds\int_{-L/2}^{L/2} f(\xi)e^{-ik \xi} d\xi,
\end{equation*}

\begin{equation*}
    \mathcal{F}^{-1}(\hat{f}(k))=f(\xi)=\ds\sum_{j\in \mathbb{Z}} \hat{f}(k) e^{ik\xi}, 
\end{equation*} 
and $\langle \,\cdot \, \rangle$  denotes the average defined by

\begin{equation*}
	\langle\Y\rangle=\ds\frac{1}{L}\int_{-L/2}^{L/2} \Y(\xi) d\xi.
\end{equation*}
 By differentiation of equation (\ref{Ything}) with respect to $\eta$ and  integration of the Cauchy-Riemann equation $X_\xi=Y_\eta$, we get }
{\color{black}
	
	\begin{equation} \label{Xeq}
		X(\xi,\eta)= \left(\frac{1+\langle\Y\rangle}{D}\right)\xi - \mathcal{F}^{-1}\left[\frac{i\cosh(k(\eta+D))}{\sinh(kD)}  \mathcal{F}(\mathbf{\Y})\right] , \quad k \neq 0.
\end{equation} 
	

The canonical depth $D$ can be fixed if we require that both canonical and physical  domains have the same length. Let $L$ and $\lambda$ be the respective lengths, thus
$$\X(\xi = L/2) - \X(\xi = -L/2)= \lambda. $$
It follows from (\ref{Xeq}) that  this  restriction  leads to the relation
\begin{equation}\label{depth}
	D=1+\langle\Y\rangle.
\end{equation}

The Laplace equation is conformally invariant. So, denoting by  $\phi(\xi,\eta)=\overline{\phi}(X(\xi,\eta),Y(\xi,\eta))$ and   $\psi(\xi,\eta)=\overline{\psi}(X(\xi,\eta),Y(\xi,\eta))$ the potential and its harmonic conjugate in the canonical coordinates, one can easily obtain that:



%

\begin{equation*}
\begin{array}{ll}
     \phi_{\xi\xi}+\phi_{\eta\eta}=0  & \hspace{10mm} \text{in} \; -D<\eta<0,\\
     \phi=\mathbf{\Phi}(\xi) & \hspace{10mm} \text{at}  \; \eta=0, \\
     \phi_\eta=0 & \hspace{10mm} \text{at} \; \eta=-D,
\end{array}
\end{equation*}
and
\begin{equation*}
\begin{array}{ll}
     \psi_{\xi\xi}+\psi_{\eta\eta}=0  & \hspace{10mm} \text{in} \; -D<\eta<0,\\
     \psi=\mathbf{\Psi}(\xi) & \hspace{10mm} \text{at}  \; \eta=0, \\
     \psi=Q & \hspace{10mm} \text{at} \; \eta=-D,
\end{array}
\end{equation*}
where $Q$ is an arbitrary constant. The formulas for $\phi(\xi,\eta)$ and  $\psi(\xi,\eta)$ can be found in similar fashion to that worked out to $X(\xi,\eta)$ and  $Y(\xi,\eta)$, which yields 
\begin{equation*}
\phi(\xi,\eta)=\mathcal{F}^{-1}\left[\frac{\cosh(k(\eta+D))}{\cosh(kD)} \mathcal{F}(\mathbf{\mathbf{\Phi}})\right], 
\end{equation*}

\begin{equation*}
	\psi(\xi,\eta)=\mathcal{F}^{-1}\left[\frac{\sinh(k(\eta+D))}{\sinh(kD)} \mathcal{F}(\mathbf{\mathbf{\Psi}})\right] - Q \frac{\eta}{D}.
\end{equation*}
Using the Cauchy-Riemman equation $\phi_\xi = \psi_\eta$ and evaluating along $\eta = 0$ we find that 

 \begin{equation}\label{Phimap}
 	\mathbf{\Phi}_\xi(\xi)=  \mathcal{F}^{-1}\left[ -i\coth(kD) \mathcal{F}_k(\mathbf{\Psi}_\xi)\right].
 \end{equation}



For simplicity, we make use of the Fourier operator $\C[\cdot]$ defined as follows: given a function $f(\xi)$, 
\begin{equation}
	\C[f(\xi)]=\C_0[f(\xi)]+\lim_{k \to 0} i\coth(kD)\hat{f}(k),
\end{equation}
where $\C_0[\;\cdot\,]=\F^{-1}\H\F[\;\cdot\,]$, with  $\H$ given by

\begin{equation}
	\H (k)=\begin{cases}
		i \coth(kD), \; k \neq 0 \\
		0, \; k=0.
	\end{cases}
\end{equation}
For the particular case of $\C[\cdot]$ evaluated at $f_\xi(\xi)$, we have that
\begin{equation}
	\C[f_\xi(\xi)]=\C_0[f_\xi(\xi)] - \frac{\hat{f}(0)}{D},
\end{equation}
With this notation, we obtain  from relations (\ref{Xeq}), (\ref{depth}) and (\ref{Phimap}) that 
\begin{equation}\label{pairs}
		\mathbf{X}_\xi=\ds 1-\C_0[\mathbf{Y}_\xi] \\
	\end{equation}

\begin{equation}\label{Phi_xi1}
			\mathbf{\Phi}_\xi=-\C_0[\mathbf{\Psi}_\xi] +\dfrac{\hat{	\mathbf{\Psi}}(0)}{D}.
	\end{equation}

 Performing straight-forward calculations  we obtain that the Kinematic condition (\ref{travelling})$_{2}$ and Bernoulli law  (\ref{travelling})$_{3}$  in canonical coordinates are given by
 
 \begin{equation}\label{Psi}
 	\Psib_\xi=c\Y_\xi-(\Omega \Y + F)\Y_\xi,
 \end{equation}
 
 \begin{equation}\label{ber2}
 	-c\ds\frac{\Phib_\xi \X_\xi +\Psib_\xi \Y_\xi}{J}+\ds\frac{1}{2J}(\Phib_\xi^2+\Psib_\xi^2)+\Y+(\Omega \Y + F)\ds\frac{\Phib_\xi \X_\xi +\Psib_\xi \Y_\xi}{J}-\Omega \Psib = 0.
 \end{equation}
Then, integrating  (\ref{Psi})  we get
\begin{equation}\label{Psi2}
	\Psib=c\Y-\left(\frac{\Omega \Y^2}{2} + F\Y\right) + M,
\end{equation}
where $M$ is an arbitrary  constant.  In order to simplify the use of the formula  (\ref{Phi_xi1})  we choose $\Psib$ so that $\hat{\mathbf{\Psi}}(0)=0$. This leads naturally to 
$$ M = \left< c\Y-\left(\frac{\Omega \Y^2}{2} + F\Y\right)\right>.  $$ 
Hence, in which follows 
\begin{equation}\label{Phi_xi}
	\mathbf{\Phi}_\xi=-\C_0[\mathbf{\Psi}_\xi].
\end{equation}

 By substituting equation (\ref{Psi2}) and (\ref{Phi_xi})   into (\ref{ber2}), then   equation (\ref{Psi}) into the resulting equation, we obtain a single equation  for the free surface  
 \begin{equation}\label{bernoulli}
 	\begin{split}
 		-\ds\frac{c^2}{2}+\ds\frac{c^2}{2J}+\Y+\ds\frac{(\C[(\Omega \Y+F)\Y_\xi])^2}{2J}-\ds\frac{\C[(\Omega \Y+F)\Y_\xi]}{J}(c-(\Omega \Y+F)\X_\xi)\\-\ds\frac{(\Omega \Y+F)^2\Y_\xi^2}{2J}-\frac{c(\Omega \Y+F)\X_\xi}{J}+Fc+\Omega\left(\frac{\Omega \Y}{2}+F\right)\Y +\Omega M=B.
 	\end{split}
 \end{equation}
 Observe that $\mathbf{X}_\xi=\ds 1-\C_0[\mathbf{Y}_\xi]$ and   $J=\mathbf{X}_\xi^2+\mathbf{Y}_\xi^2$, so    this is an equation whose unknowns are $\Y(\xi)$, $c$, $D$ and $B$.} It is the aim of the next section to describe an  approach for finding solitary-type solutions numerically.

\subsection{Numerical method}\label{steadymethod}

Up to this point, we have transformed the free boundary problem  (\ref{travelling})  into an algebraic system of  two equations (  (\ref{depth}) and (\ref{bernoulli}) )  and four unknowns $\Y(\xi)$, $c$, $D$ and $B$. 	In order to get a system  that can  be handled by Newton's method, we add two extra equations.

We  fix the amplitude $A$ of the wave through 
\begin{equation}\label{amplitude}
	Y(0)-Y(-L/2)=A,
\end{equation}
and based on the limit (\ref{limite}) we impose that 
\begin{equation} \label{limite2}
Y(-L/2)= 0.
\end{equation}

Consider a discrete version of  equations  \ref{depth}, \ref{bernoulli}, \ref{amplitude} and \ref{limite2} as follows. Let us take an evenly spaced grid in the $\xi$ axis in the canonical domain, say 
\begin{equation} \label{grid}
    \xi_j=-L/2+(j-1)\Delta\xi, \hs{10pt} j=1,...,N, \; \hs{10pt} \text{where } \Delta\xi=L/N, 
\end{equation}
with $N$ even. We impose symmetry about $\xi=0$ so that $Y_j=Y_{N-j+2}$, where $Y_j=\Y(\xi_j)$.
	Fixing $\Omega$ and $F$, we have  $N/2+4$ unknowns: $Y_1, \cdots, Y_{N/2+1}$, $c$, $D$ and $B$. We satisfy 	 equation (\ref{bernoulli}) at the grid points (\ref{grid}). The  Fourier modes are computed by the Fast Fourier Transform (FFT) and derivatives in the $\xi$-variable are performed spectrally (\cite{Trefethen}).   This results in $N/2+1$ equations $$\mathcal{G}_j (Y_1, \cdots, Y_{N/2+1}, c, D,B)  = 0 \quad j = 1, \cdots, N/2+1$$.  Equation (\ref{depth}) is discretized using the trapezoidal rule, which leads to the equation
	\begin{equation*}
	\mathcal{G}_{N/2+2} (Y_1, \cdots, Y_{N/2+1}, c, D,B)= \frac{Y_1+Y_{N/2+1}}{2}+\sum_{j=2}^{N/2}Y_j+1-D=0.
	\end{equation*}
	Finally, we satisfy (\ref{amplitude}) and (\ref{limite2}), resulting in a system of the $N/2+4$ equations and $N/2+4$ unknowns, 
	$$\mathcal{G}_{N/2+3} (Y_1, \cdots, Y_{N/2+1}, c, D,B)= Y_{N/2+1}-Y_1 - A = 0,$$
	$$\mathcal{G}_{N/2+4} (Y_1, \cdots, Y_{N/2+1}, c, D,B)= Y_1 = 0.$$

	The system is solved by Newton's method, where our initial guess is taken to be the well known solitary wave solution for the classical (irrotational) Koterweg-de Vries equation, that is

\begin{align*}
    \Y(\xi)=A_0\sech^2\left(\sqrt{3A_0/4}\xi\right), \; c=1+\frac{A_0}{2}, \;
\end{align*}
where $A_0$ is chosen small ($0.02$ was used). From there, the idea is to make use of the continuation technique in $A$ and $\Omega$, where the prior converged solution in fed as initial guess to a new solution. The Jacobian matrix of the system is computed by finite difference and the stopping criterion  for the Newton's  method is 
	$$ \frac{\sum_{j=1}^{N/2+4}|\mathcal{G}_j|}{N/2+4} < 10^{-10}.$$

In all experiments performed we used  $L = 1500$. This is important to make sure that the method indeed converges to a solitary-type solution instead of something else, such as known periodic solutions which can be captured by a similar method, presented in \cite{Ribeiro Jr}. Finally, for each choice of $\Omega$, the Froude number used was, in all the examples that will be presented later, $F= {\Omega}/{2}$. This implies canceling the average mass flow of the stream $\U_0=(\Omega Y + F,0)$.

\section{Time-dependent solitary waves}\label{time}

We now shift the attention to a brief discussion on time-dependent Euler equations (\ref{td}) and their solutions as it is one of the main concerns of this work to develop a technical comparison between steady and non-steady frameworks. This will be detailed further, in section \ref{evol sec}.

{\color{black}
Note that there is no loss of generality in assuming that the constant of Bernoulli  $B(t)$ in the boundary condition  (\ref{td})$_3$ is equal to zero. Having said this, the numerical method that we will use to study the time dependent Euler equations is a simpler version of the one presented by  \cite{Marcelo-Paul-Andre}. These authors investigated waves generated by current-topography interaction in Euler equations framework through a time-dependent conformal mapping. In our work we consider the flat bottom version of their method which is summarized here. }

The now time-dependent mapping is given by
\begin{equation*}
z(\xi,\eta,t) = x(\xi,\eta,t)+iy(\xi,\eta,t),
\end{equation*}
which flattens the free surface and maps a strip of width $D(t)$ onto the fluid domain and satisfies the boundary conditions 
\begin{equation*}
y(\xi,0,t)=\zeta(x(\xi,0,t),t) \;\ \mbox{and} \;\ y(\xi,-D(t),t)=-1.
\end{equation*}


Let $\phi(\xi,\eta,t)=\bar{\phi}(x(\xi,\eta,t),y(\xi,\eta,t),t)$ and $\psi(\xi,\eta,t)=\bar{\psi}(x(\xi,\eta,t),y(\xi,\eta,t),t)$ be its harmonic conjugate, and  denote by
$\mathbf{\Phi}(\xi,t)$ and $\mathbf{\Psi}(\xi,t)$ their traces along $\eta=0$ respectively. Considering $\mathbf{X}(\xi,t)$ and $\mathbf{Y}(\xi,t)$ as the horizontal and vertical coordinates at $\eta=0$, Kinematic and Bernoulli conditions $(\ref{td})_{2,3}$ are read as
\begin{align}\label{eulerconforme}
\begin{split}
& \mathbf{Y}_{t} =\mathbf{Y}_{\xi}\mathcal{C}\bigg[\frac{\mathbf{\Theta}_{\xi}}{J}\bigg] 
-\mathbf{X}_{\xi}\frac{\mathbf{\Theta}_{\xi}}{J}, \\
& \mathbf{\Phi}_{t} = - \mathbf{Y} - \frac{1}{2J}
(\mathbf{\Phi}_{\xi}^{2}-\mathbf{\Psi}_{\xi}^{2}) +\mathbf{\Phi}_{\xi}\mathcal{C}\bigg[\frac{\mathbf{\Theta}_{\xi}}{J}\bigg] 
- \frac{1}{J}(F+\Omega Y)\mathbf{X}_{\xi}\mathbf{\Phi}_{\xi},
\end{split}
\end{align}
where $\mathbf{\Theta}_{\xi}(\xi,t)=\mathbf{\Psi}_{\xi}+F\mathbf{Y}_{\xi}+\Omega \mathbf{Y}\mathbf{Y_\xi}$, $J=\mathbf{X}_{\xi}^2+\mathbf{Y}_{\xi}^2$ is the Jacobian of the conformal mapping evaluated at $\eta=0$, 
\begin{align*}
\begin{split}
& \mathbf{X}_{\xi} = \ds\frac{1}{D}-\mathcal{C}\big[\mathbf{Y}_{\xi}\big], \\
& \mathbf{\Phi}_{\xi} = -\mathcal{C}\big[\mathbf{\Psi}_{\xi}\big], \\
\end{split}
\end{align*}
and $\mathcal{C}$ is the operator introduced in the previous section. As before, we define the now time-dependent canonical depth by 

$$D(t) = \left<\Y(\xi,t)\right>+1,$$
so that wavelength is preserved throughout the transformation. Details on the method can be found in \cite{Marcelo-Paul-Andre}. 

The wave dynamic is found integrating in time the family of ordinary differential equations (\ref{eulerconforme}) through the fourth-order Runge-Kutta method.  Besides, the canonical domain is discretized as mentioned in the steady wave case (see equation \ref{grid}) with all Fourier transforms approximated by the FFT.  

\section{Results} \label{Sec_Results}

\subsection{Steady waves}

\begin{figure}[H]
    \centering
    \includegraphics{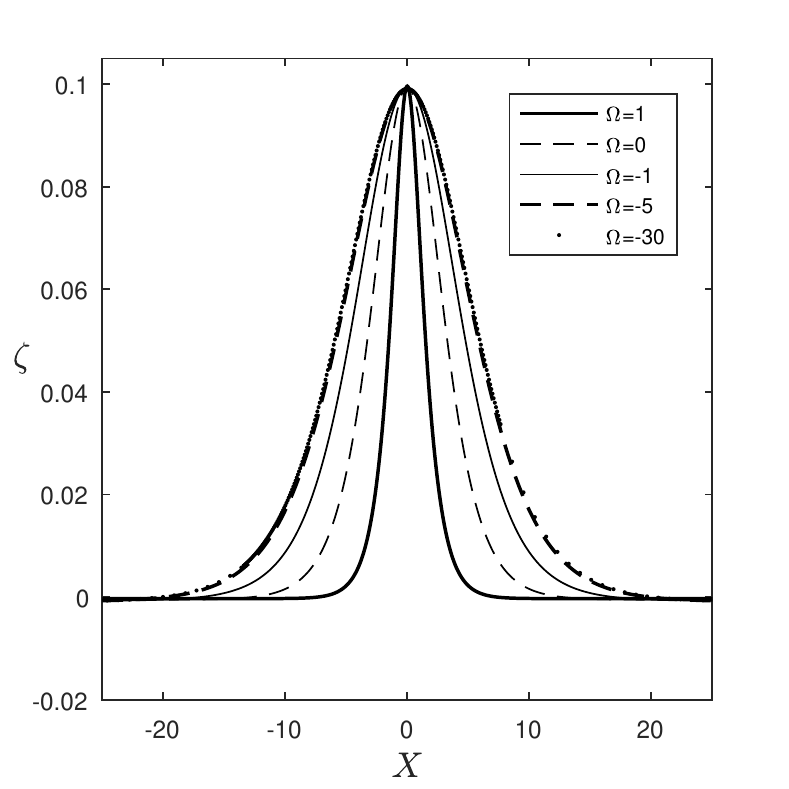}
    \includegraphics{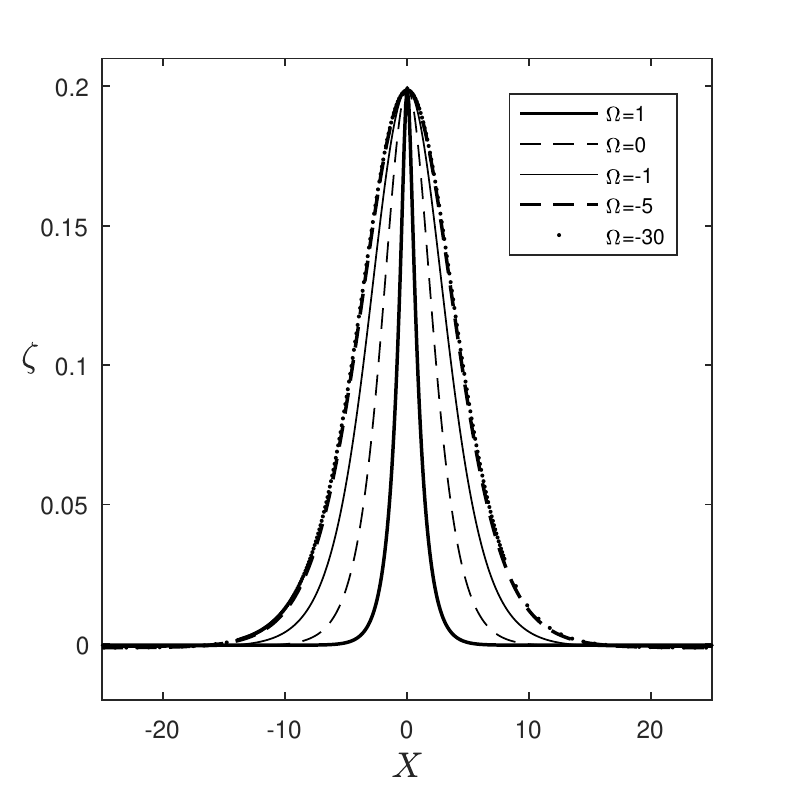}
    \caption{Wave profiles with amplitudes $A=0.1$ (left) and $A=0.2$ (right). Although the computational domain utilized was 1500, for visualization purposes the plotting window size was chosen to be 50 length units wide.}
    \label{profiles}
\end{figure}

It is well known in the literature  that the crests of the waves become rounder as  $\Omega$ decrease. This has been shown by \cite{Da silva e Peregrini}, \cite{Vanden-Broeck96}, \cite{Ko Europe}, \cite{Ribeiro Jr} and  \cite{DyachenkoHur2} for periodic travelling waves and by \cite{Vanden-Broeck94} for solitary waves. Figure \ref{profiles} displays various wave profiles for different vorticity values. As can be seen, the numerical method captures these well known characteristics about waves with vorticity: more rounded or cuspidate profiles depending on the $\Omega$ signal. Although the computational domain used was equal to 1500, for visualization purposes the plot window was chosen to be 50 units long.





\begin{figure}[H]
    \centering
    \includegraphics{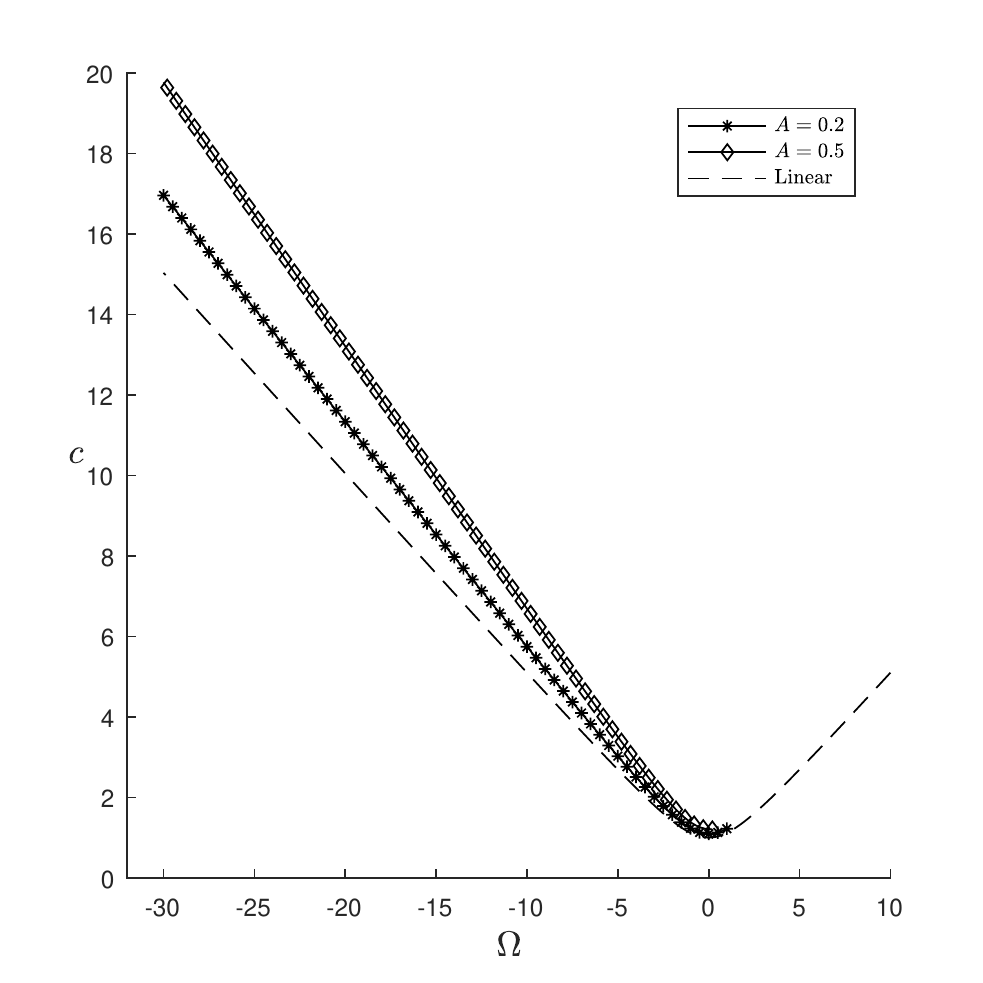}
    \caption{Velocity - vorticity relation}
    \label{c_gamma}
\end{figure}

Furthermore, vorticity also has a straight-forward and expected effect in the velocity of the constructed waves: greater vorticity implies greater velocity across the amplitude spectrum, a trend that matches with the well-known dispersion relation from linear theory, as depicted in Figure \ref{c_gamma}.

The closed formula for the velocity shown in dashed lines in Figure \ref{c_gamma} is given by

\begin{equation*}
    c_{lin}=F-\ds\frac{\Omega}{2}+\sqrt{\ds\frac{\Omega^2}{4}+1}.
\end{equation*}




In what follows, we can see how the velocities are influenced by increasing the amplitude while keeping vorticity fixed. For low amplitudes, it is expected that waves constructed by the presented method show similar behaviour to the $\sech^2$-type solution of the KdV equation. Regarding the analysis a KdV model in the presence of vorticity we refer to \cite{Guan}, as the formulation presented there was used here to fix exact solutions of a reduced model as reference for comparison with our numerical solutions.


\begin{figure}[H]
    \centering
    \includegraphics{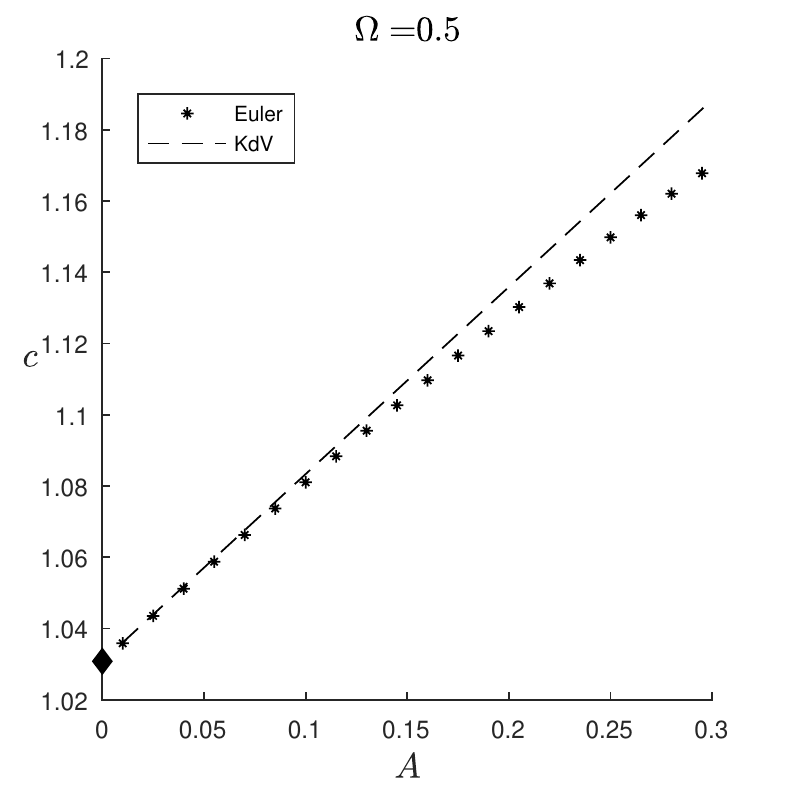}
    \includegraphics{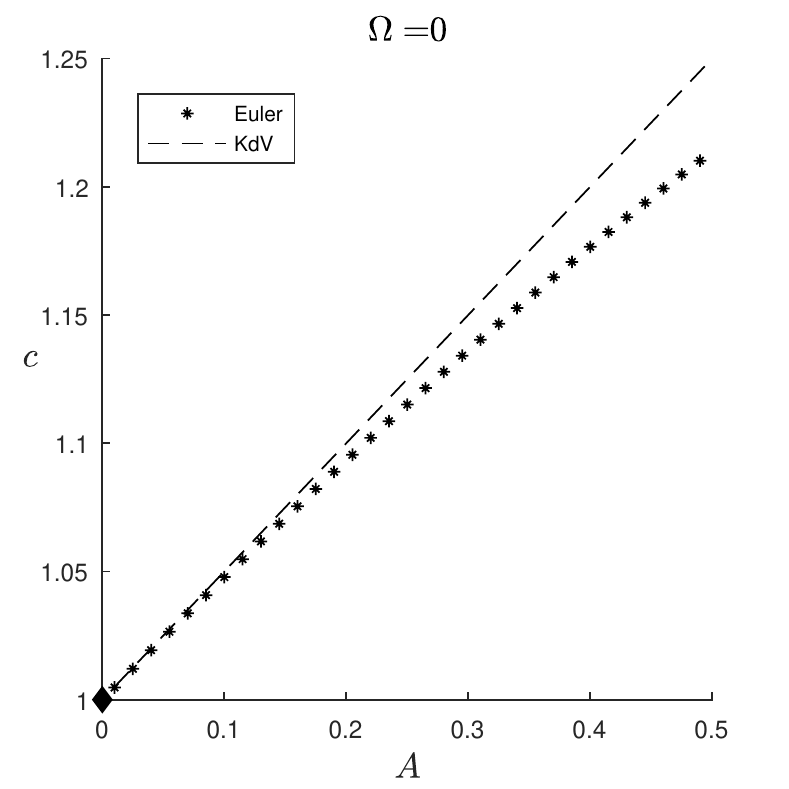}
    \includegraphics{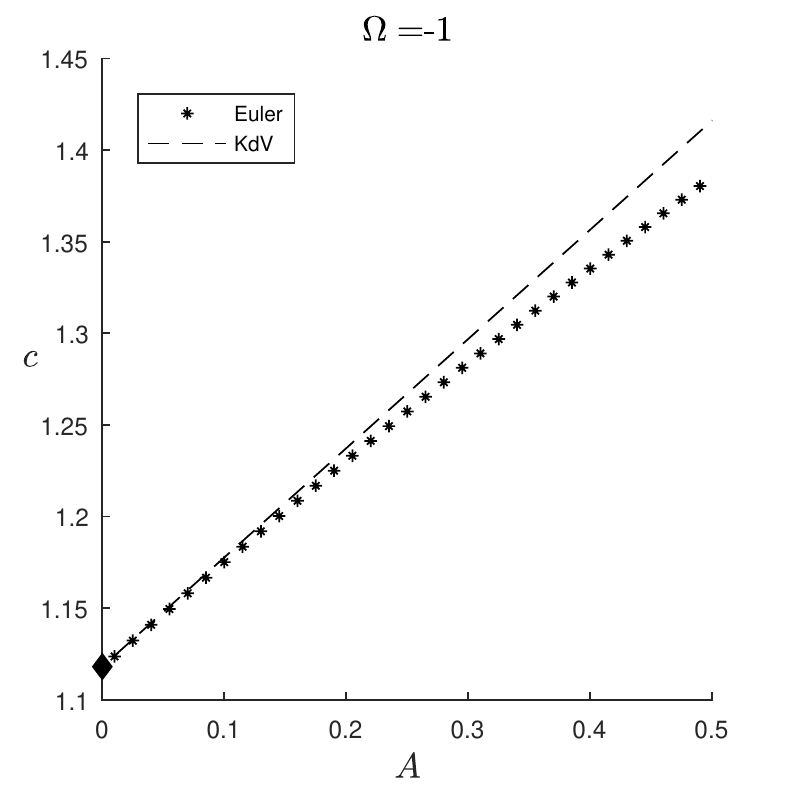}
    \includegraphics{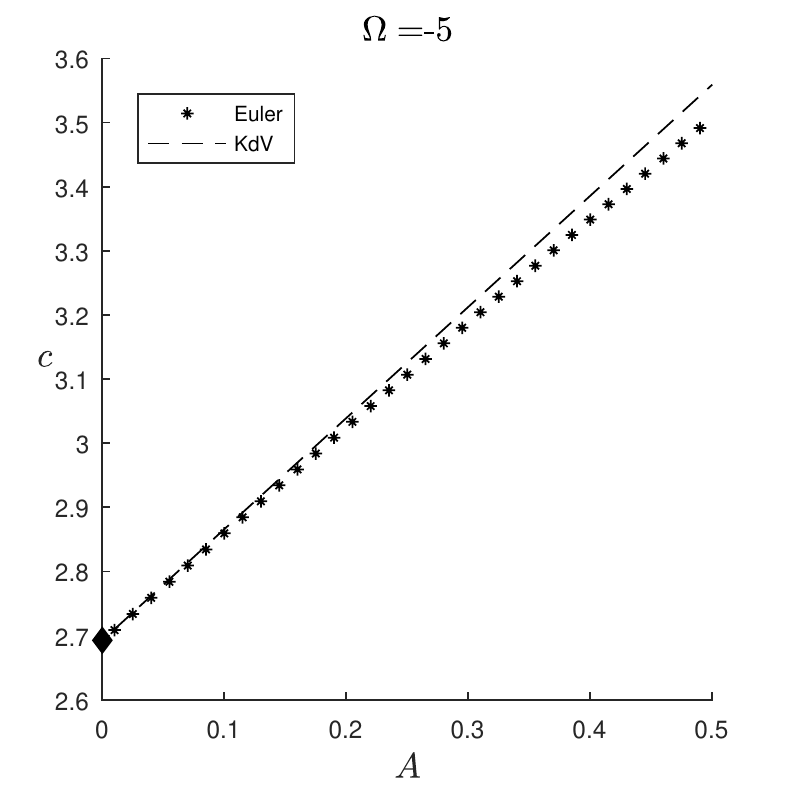}
    \caption{Velocity - amplitude relation, for different choices of vorticity and comparison to the linear relation from KdV}
    \label{KdV vel}
\end{figure}

The dashed line means the rotational KdV velocity solution after scaling to the Euler regime, specifically
\begin{equation}\label{cKdV}
	c_{KdV}=\tilde{c}+\epsilon\tilde{v}+F,
\end{equation}
where

\begin{equation}
    \begin{cases}
    \tilde{c}=\ds\frac{-\Omega+\sqrt{\Omega^2+4}}{2},\\[10pt]
    \delta = \sqrt{\ds\frac{(\Omega^2+3)}{4\tilde{c}^2}}, \\[10pt]
    \tilde{v}=\ds\frac{4\delta^2\tilde{c}}{3(2\tilde{c}++\Omega)},
    \end{cases}
\end{equation}
and $\epsilon$ is the non-linearity parameter.

For a given choice of parameters $\Omega$ and $A$ and in a certain sense, Figure \ref{KdV vel} can indicate the distance between our solutions to the analytical solution found in KdV regime. As expected we see very close velocities whenever $A$ is small but the overall pattern of velocity/amplitude relation in the case of Euler solutions present a clear deviation from the linear distribution found in KdV. In particular, around $A=0.15$, $A=0.2$ we see a slight takeoff from the Euler regime in comparison to the KdV, while it is interesting to observe that the general aspect of this ``takeoff curve'' is overall maintained unchanged when we vary vorticity choices. Finally, we remind that, at best, there will always be slight deviations on numerical output depending on the number of mesh points. The confidence in the results presented throughout this section and the next is greatly based on a resolution study which is detailed in Appendix \ref{reso}.

\subsection{Evolution of solitary waves}\label{evol sec}

In this section, we make use of the numerical evolution method synthesized in section \ref{time} to analyze the actual dynamics of our constructed waves, that is, to check whether the solutions given by our steady waves numerical method are indeed steady/travelling solutions. 


Assume the solitary wave, of amplitude $A$, solution of the flow of vorticity $\Omega_0$ and Froude number $F_0$ ($\Omega_0/2$) is computed numerically. Let us denote by $\Y_0$ its profile and $c_0$ its velocity. Now we are interested to solve system (\ref{eulerconforme}) with parameters $\Omega=\Omega_0$ and $F=F_0-c_0$. With this choice, we are including a countercurrent which cancels out precisely the wave velocity, leaving a stationary wave as the solution, if $c_0$ was to be its original speed. Also, a numerical calculation of $\Phib_0$ from $\Y_0$ is necessary, but this can be achieved by

\begin{equation}
    \Phib_0=\mathcal{F}_{k\neq0}^{-1} \left[ -i\ds\frac{1}{k} \hat{\Phib}_\xi(k) \right],
\end{equation}
taken $(\ref{pairs})_2$ and (\ref{Psi}) into account. Initial data can now be fixed as


\begin{equation}
\begin{cases}
    \Y(\xi,0)=\Y_0, \\
    \Phib(\xi,0)=\Phib_0.
\end{cases}
\end{equation}

Letting $f(t)$ denote the vector with entries $(f(\xi_j,t))$, $j=1,...,N/2+1$, an error function is defined by:

\begin{equation}
    E(t)=\ds\frac{||\Y(t)-\Y_0||_{\infty}}{||\Y_0||_{\infty}}.
\end{equation}

Essentially, we want to investigate how the vorticity, as well as amplitude, of a given Euler solution, affects the numerical stability of the wave. For the criterion, we make use of the empirical quantity called the \textit{effective wavelength}, denoted by $EW$, which is the length of the interval that contains almost all of the energy of the wave. The test is then whether the wave would be able to travel (without changing shape or velocity) a distance which is at least 25 times its effective wavelength, were us to withdraw the countercurrent. All we need to do is see if the wave holds still and stable up to time $t=\lfloor 25 \times EW/c \rfloor$ or longer. The measuring of shape and velocity preservation during evolution will be then calculated by the error function $E(t)$. As for the Runge-Kutta fourth-order evolution, the time step choice was $\Delta t=10^{-2}$. Here, the time grid used is given by $t_j=(j-1)\Delta t, \; j=1,...,Nt$ with $Nt=2000/\Delta t$.

When selecting output data from our method to feed as initial data for the evolution code, we chose $\Delta\xi=0.1831$ as the reference grid interval for this experiment. As typically, the refinement of space and/or time grid may influence results specially in limiting cases, mainly critical amplitude/vorticity waves, which may lie on or close to the boundaries of the solution space. These seemingly arbitrary choices are actually suggested by a mixture of empirical analysis and a rigorous resolution study, which can be found in Appendix \ref{reso}.

Table \ref{evolution} synthesizes the core of the results of this section.


\begin{table}[H]
\centering
\begin{tabular}{|c|c|c|c|}
\hline
$A$ & $\Omega$ & $T$ & $E(t=T)$ \\ \hline
\multirow{6}{*}{$0.1$} & $1$ & $2000$ & $3.92 \times 10^{-9}$ \\ \cline{2-4} 
 & $0.5$ & $2000$ & $4.01 \times 10^{-7}$ \\ \cline{2-4} 
 & $0$ & $2000$ & $2.10 \times 10^{-8}$ \\ \cline{2-4} 
 & $-0.5$ & $2000$ & $3.47 \times 10^{-9}$ \\ \cline{2-4} 
 & $-1$ & $2000$ & $8.36 \times 10^{-12}$ \\ \cline{2-4} 
 & $-5$ & $74.95$ & $0.0415$ \\ \hline
\multirow{6}{*}{$0.2$} & $1$ & $2000$ & $0.1167$ \\ \cline{2-4} 
 & $0.5$ & $2000$ & $3.21 \times 10^{-7}$  \\ \cline{2-4} 
 & $0$ & $2000$ & $1.14 \times 10^{-12}$  \\ \cline{2-4} 
 & $-0.5$ & $2000$ & $1.44 \times 10^{-9}$ \\ \cline{2-4} 
 & $-1$ & $2000$ &  $5.62 \times 10^{-5}$ \\ \cline{2-4} 
 & $-5$ & $6.7900$ & $0.0045$ \\ \hline
\multirow{6}{*}{$0.3$} & $1$ & * & * \\ \cline{2-4} 
 & $0.5$ & $2000$ & $0.0016$ \\ \cline{2-4} 
 & $0$ & $2000$ & $4.67 \times 10^{-12}$ \\ \cline{2-4} 
 & $-0.5$ & $58.51$ & $0.0855$ \\ \cline{2-4} 
 & $-1$ & $21.77$ & $0.0410$ \\ \cline{2-4} 
 & $-5$ & $4.11$ & $0.0700$ \\ \hline
\end{tabular}
\caption{Stability study: solutions with different parameters present distinct stability levels and may blow up prior to time $t=2000$. (The row marked with ``*'' means the method did not converge)}
\label{evolution}
\end{table}

 For this experiment, $T=\min\{2000, t^*\}$, where $t^*=\max\{t_j; j=1,...,N_t \; ; \;  E(t_j) < 1\}$. The reference time $t=2000$ was chosen as it guarantees that all of the selected waves travel at least 25 times their effective wavelength. 
 
We note that, in general, waves are unstable for values of $\Omega$ which are large in modulus. For $\Omega>0$, this instability is somewhat to be expected, as the solution loses smoothness as $\Omega$ grows (as seen in the figure \ref{profiles} the wave tends to form cusps in the neighborhood of $X=0 $ to $\Omega \gg 0$.) On the other hand, for $\Omega<0$, although the solutions are smooth for values of $\Omega$ whose modulus is arbitrarily large, the waves are unstable beyond a certain critical value. This indicates that although we can calculate traveling solitary waves for any negative value of $\Omega$, existence of these waves for $|\Omega| \gg1$ is just theoretical -- in the sense that they are traveling solutions of Euler's equations that in practice do not travel.
 
Furthermore, the data in table \ref{evolution} indicate that the wave amplitude parameter is directly correlated to the emergence of instability. In other words, setting $\Omega \neq0$, increase in amplitude leads to an instability in the solution. This happens even when $|\Omega|$ is small, as indicated by the third column. It can be highlighted that the case $\Omega=0$ is stable for all amplitudes presented in the table, which indicates that the stability of rotational waves tends to be more sensitive to amplitude increases.

\section{Conclusion}

We were able to construct an alternative computational via through which solitary-type solutions to Euler equations can be found, in the presence of constant vorticity. The derivation of this method, which is based on successful approaches to similar problems presented in previous works, is thoroughly explained from the governing equations themselves as a starting point. It mainly consists of the conformal mapping technique, the spectral derivative approach to PDEs and Newton's method coupled with a continuation scheme on the relevant characteristic parameters: amplitude ($A$) and vorticity ($-\Omega$). Effects of these parameters on the solutions are then visually presented and rigorously discussed, also maintaining linear theory as well as the classic Kortweg-de Vries solitary solution (weakly nonlinear theory) as reference for comparison and validation. Finally, we proceed to put the method to test when we take our constructed solutions and feed them as initial data to the evolution code for time-dependent method for Euler equations, making it possible to get an insight on the stability of the waves found and an idea of the limits of the solution space for this problem. Inclusion of aspects which were taken trivially in the assumptions of this problem - pressure, topography, etc. - might be considered as natural potential progression from the present work, with the idea of investigating the combination of effects in a more general way, as well as a complete discussion on trajectories. Another seemingly natural extension of the work exposed here could include the consideration of more general, non-constant, vorticity distribution when choosing the shear current flow.

\section*{Acknowledgements}

The author E.M.C. is grateful for the financial support provided by CAPES Foundation (Coordination for the Improvement of Higher Education Personnel) during part of the development of this work.

\bibliographystyle{abbrv}

\begin{thebibliography}{99}
	
	\bibitem[Toland et al. (1982)]{Toland1}
	\textsc{Amick, C., Fraenkel, L., Toland, J.} 1982
	On the Stokes conjecture for the wave of extreme form. {\it Acta Mathematica} \textbf{148}, 193-214.
	
	\bibitem[Toland et al. ($1981^1$)]{Toland3}
	\textsc{Amick, C., Toland, J.} 1981
	On periodic water waves and their convergence to solitary waves in the long-wave limit.  {\it Philos. Trans. R. Soc. London Ser. A Math. Phys. Eng. Sci.  } \textbf{303}, 633-669.
	
	\bibitem[Toland et al. ($1981^2$)]{Toland2}
	\textsc{Amick, C., Toland, J.} 1981
	On solitary water waves of finite amplitude. {\it Arch. Ration. Mech. Anal.} \textbf{76}, 9-95.
	
	\bibitem[Beale (1977)]{Beale}
	\textsc{Beale, J.} 1997
	The existence of solitary water waves. {\it Comm. Pure Appl. Math.} {\bf 30}, 373-389.
	
	\bibitem[Benjamin (1962)]{Benjamin2}
	\textsc{Benjamin, T. B.} 1962
	The solitary wave on a stream with an arbitrary distribution of vorticity.
	{\it J. Fluid Mech.} \textbf{12}, 97-116.
	
	
	
	

	
	     \bibitem[Choi (2003)]{Choi}
	  \textsc{Choi W.} 2003
{Strongly nonlinear long gravity waves in uniform shear flows}
{\it Physical Review E}. \textbf{68}, 026305.

    \bibitem[Choi (2009)]{Choi2}
	  \textsc{Choi W.} 2009
{Nonlinear surface waves interacting with a linear shear current}
{\it Mathematics and Computers in Simulation}. \textbf{80}, 29-36.
	
	
	
	\bibitem[Dyachenko \& Hur (2019)]{DyachenkoHur1}
	\textsc{Dyachenko, S., Hur, V.} 2019 
	Stokes waves with constant vorticity: folds, gaps and fluid bubbles.
	{\it J. Fluid Mech.} \textbf{878}, 502-521.
	
\bibitem[Dyachenko \& Hur. (2019)]{DyachenkoHur2}
\textsc{Dyachenko, S., Hur, V.} 2019 
{Stokes waves with constant vorticity: I. Numerical
	computation }
{\it Stud Appl Math}, {\bf 142}, 162-189.	
	
	
	
	
\bibitem[Dyachenko et al. (1996)]{Dyachenko}
	  \textsc{Dyachenko AL, Zakharov VE, Kuznetsov EA.} 1996 
{Nonlinear dynamics of the free surface of an ideal fluid.}
{\it  Plasma Phys.}22:916-928. 

	\bibitem[Flamarion et al.  (2019)]{Marcelo-Paul-Andre}
	\textsc{Flamarion, M., Milewski, P., Nachbin A.} 2019 
	Rotational waves generated by current-topography interaction.
	{\it Stud Appl Math}, {\bf 142}, 433-464.

    \bibitem[Francius \& Kharif (2017)]{Kharif}	
	\textsc{Francius, M.; Kharif, C.} 2017
	Two-dimensional stability of finite-amplitude gravity waves on 	water of finite depth with constant vorticity. {\it J. Fluid Mech.}  \textbf{830}, 631-659.

    \bibitem[Freeman \& Johnson (1970)]{Freeman}
	\textsc{Freeman, N. G., Johnson, R. S.} 1970
	Shallow water waves on shear flows.  {\it J. Fluid Mech.} \textbf{42}, 401-409.

    \bibitem[Friedrichs \& Hyers (1954)]{Friedrichs}
	\textsc{Friedrichs, K., Hyers, D.} 1954
	The existence of solitary waves. 
	{\it  Comm. Pure Appl. Math.} {\bf 7}, 517-550.

	\bibitem[Guan (2020)]{Guan}
	\textsc{Guan, X.} 2020 
	Particle trajectories under interactions between solitary waves and a linear shear current.
	{\it Theoretical \& Applied Mechanics Letters}. {\bf 10}, 125-131.

    \bibitem[Hur (2008)]{Hur}
	\textsc{Hur, V. M.} 2008
	Exact solitary water waves with vorticity. {\it Arch. Ration. Mech. Anal.} \textbf{188}, 213-244.

    \bibitem[Johnson (1986)]{Johnson}
	\textsc{Johnson, R. S.} 1986
	On the nonlinear critical layer below a nonlinear unsteady surface wave. {\it Journal of Fluid Mechanics}, \textbf{167}, (327-351).


    \bibitem[Ko \& Strauss (2008)]{Ko Europe}
	  \textsc{Ko J, Strauss W.} 2008
{Large-amplitude steady rotational water waves}
{\it European Journal of Mechanics}. \textbf{27}, 96-109.


    \bibitem[Kozlov et al. (2020)]{Kozlov}
	\textsc{Kozlov, V.; Kuznetsov, N; Lokharu, E.} 2020
	Solitary waves on constant vorticity flows with an interior stagnation point. {\it J. Fluid Mech.}, {\bf 904}.


    \bibitem[Lavrentiev (1954)]{Lavrentiev}
	\textsc{Lavrentiev, M.} 1954
	On the theory of long waves I A contribution to the theory of long waves II. {\it Amer. Math. Soc. Transl.} {\bf Ser. 2 102}, 3-50.

 \bibitem[Milewski et al. (2010)]{Milewski}
\textsc{Milewski, P. A., Vanden-Broech, J.-M. \& Wang, Z.} 2010
{Dynamics of steep two-dimensional gravity-capillary solitary waves}
{\it J. Fluid Mech.}. \textbf{664}, 466-477.
	
		  \bibitem[Ribeiro Jr et al (2017) ]{Ribeiro Jr}
	\textsc{Ribeiro Jr., R., Milewski, P. A. \& Nachbin, A.} 2017
	Flow structure beneath rotational water waves with stagnation points.
	{\it J Fluid Mech}. \textbf{812}, 792-814.
	
      \bibitem[Teles da Silva \& Peregrini (1988) ]{Da silva e Peregrini}
	\textsc{Teles da Silva, A.F.,  Peregrini, D.H.} 1984
	Steep, steady surface waves on water of finite depth with constant vorticity.
	{\it J Fluid Mech}. \textbf{195}, 281-302.
	
	\bibitem[Ter-Krikorov (1960)]{Ter-Krikorov}
	\textsc{Ter-Krikorov, A. M.} 1960
	The existence of periodic waves which degenerate into a solitary wave. 
	{\it J. Appl. Math. Mech.} {\bf 24}, 930-949.
	
	
	
	
	\bibitem[Trefethen (2001)]{Trefethen}
    \textsc{Trefethen, L. N.} 2001
    Spectral Methods in MATLAB.
    {\it Philadelphia: SIAM.}
	
	 \bibitem[Vanden-Broeck (1996) ]{Vanden-Broeck96}
	\textsc{Vanden-Broeck, J.-M.} 1996
    Periodic waves with constant vorticity in water of infinite depth.
    {\it IMA J. Appl. Maths }. \textbf{56}, 207-217.

\bibitem[Vanden-Broeck (1994) ]{Vanden-Broeck94}
\textsc{Vanden-Broeck, J.-M.} 
1994 Steep solitary waves in water of finite depth with constant vorticity.
{\it J. Fluid Mech }. \textbf{274}, 339-348.

    \bibitem[Wheeler (2013)]{Wheeler}
	\textsc{Wheeler, M.} 2013
	Large-amplitude solitary water waves with vorticity. {\it SIAM J. Math. Anal.} \textbf{45 (5)}, 2937-2994.
	

	
	
	

\end{thebibliography}

\appendix 

\section{Resolution study}\label{reso}

In what follows we show that the method is  independent of the grid size by calculating the distance between outputs for different choices of $\Delta \xi$. These experiments were performed for waves with amplitude $A=0.2$.  We take the reference grid  as  $\Delta \xi^*=0.0458$, the  finest resolution computed.
\begin{table}[H]
\begin{center}
\begin{tabular}{ c|c|c|c } 
 \hline
    $\Omega$ & $\Delta \xi$ & $||\zeta_{\Delta \xi} - \zeta^* ||_2/||\zeta^*||_2$ & $|c_{\Delta \xi} - c^*|/|c^*| $ \\ \hline \hline
     $0$ & $0.0916$ & $1.6\times10^{-10}$ & $2.5\times10^{-12}$ \\ \hline  
     $0$ & $0.1831$ & $4.6\times10^{-10}$ & $7.4\times10^{-12}$ \\ \hline 
     $0$ & $0.3662$ & $6.7\times10^{-8}$ & $5.6\times10^{-10}$ \\ \hline 
     $0$ & $0.7324$ & $1.2\times10^{-4}$ & $5.2\times10^{-6}$ \\ \hline \hline 
    
    $-1$ & $0.0916$ & $5.4\times10^{-13}$ & $2.1\times10^{-14}$ \\ \hline  
    $-1$ & $0.1831$ & $5.7\times10^{-13}$ & $1.0\times10^{-14}$ \\ \hline 
    $-1$ & $0.3662$ & $1.5\times10^{-12}$ & $1.3\times10^{-14}$ \\ \hline 
    $-1$ & $0.7324$ & $1.2\times10^{-07}$ & $1.8\times10^{-09}$ \\ \hline \hline
    $1$ & $0.0916$ & $6.8\times10^{-05}$ & $1.0\times10^{-06}$ \\ \hline  
    $1$ & $0.1831$ & $0.0027$ & $1.7\times10^{-04}$ \\ \hline 
    $1$ & $0.3662$ & $0.0477$ & $0.0015$ \\ \hline 
    $1$ & $0.7324$ & $0.1164$ & $0.0102$  \\ 
\end{tabular}
\end{center}
\caption{Resolution study for waves of amplitude $A=0.2$.}
\label{tabela2}
\end{table}

 In table \ref{tabela2}, we denote by $\zeta_{\Delta \xi}$  the wave profile and by $c_{\Delta \xi}$    the wave speed  obtained from the Newton's method using a grid  with size $\Delta \xi$. In addition, we consider as  $\zeta^*$  
and  $c^*$ the wave profile and its speed computed in the finest grid. 
These experiments were performed for waves with amplitude $A = 0.2$. 
Note that for  $\Omega = 1$ the numerical scheme requires more resolution for approximating  the solution with more accuracy. This  can be explained by a combination of two factors: i) the emergence of cusps; ii) the issue of crowding phenomenon present in conformal mappings. For this reason, finer grids are necessary to accurately  compute waves in presence of currents where  $\Omega$ is positive.

\end{document}